\documentclass[pra, reprint, twocolumn, superscriptaddress, amsmath, amssymb, aps]{revtex4-2}

\usepackage[english]{babel}
\usepackage{graphicx}
\graphicspath{{./Images/},{./ImagesAppendix/},{./images/},{./imagesAppendix/}}

\usepackage{amsmath, amssymb, amsfonts}
\usepackage{physics}
\usepackage{bm}
\usepackage{braket}
\usepackage{array}
\usepackage{multirow}
\usepackage{float}
\usepackage{commath}
\usepackage{verbatim}
\usepackage{color}
\usepackage{enumitem}
\usepackage{array}   
\usepackage{dsfont}
\usepackage{mathtools}
\usepackage{soul}
\usepackage[colorlinks=true, citecolor=darkBlue, linkcolor=darkBlue, urlcolor=blue]{hyperref}
\usepackage{amssymb}
\usepackage{tabularx}
\usepackage{booktabs}
\definecolor{darkBlue}{rgb}
{0.08, 0.13, 0.4}

\usepackage{cancel}

\usepackage{tikz}
\usetikzlibrary{positioning, arrows.meta}
\usepackage{pgfplots}

\definecolor{qred}{RGB}{210,40,50}
\definecolor{qredpurple}{RGB}{185,55,120}
\definecolor{qpurple}{RGB}{145,65,170}
\definecolor{qbluepurple}{RGB}{90,75,180}
\definecolor{qblue}{RGB}{35,90,190}

\definecolor{THc}{rgb}{0.9,0.3,0.2}

\newcommand{\canc}[1]{}

\begin{document}

\title{Coherence-driven quantum active systems}
\author{F. Perciavalle}
\affiliation{Dipartimento di Fisica, Universit\`a della Calabria, 87036 Arcavacata di Rende (CS), Italy}
\affiliation{INFN, Sezione LNF, gruppo collegato di Cosenza}

\author{G. Jacucci}
\affiliation{Dipartimento di Fisica, Universit\`a della Calabria, 87036 Arcavacata di Rende (CS), Italy}
\affiliation{INFN, Sezione LNF, gruppo collegato di Cosenza}
\affiliation{CNR Nanotec, Istituto di Nanotecnologia S.S. Cosenza, Rende (CS)}

\author{F. Plastina}
\affiliation{Dipartimento di Fisica, Universit\`a della Calabria, 87036 Arcavacata di Rende (CS), Italy}
\affiliation{INFN, Sezione LNF, gruppo collegato di Cosenza}

\author{N. Lo Gullo}
\affiliation{Dipartimento di Fisica, Universit\`a della Calabria, 87036 Arcavacata di Rende (CS), Italy}
\affiliation{INFN, Sezione LNF, gruppo collegato di Cosenza}

\date{\today}

\begin{abstract}
Starting from a classical diffusive process, the introduction of coherent hopping gives rise to active-like motion, with quantum coherence acting as a ``fuel'' for activity. An exact hierarchy of equations of motion connects particle populations, currents, and coherences at progressively larger distances. This hierarchy provides a natural framework for characterizing quantum active dynamics beyond particle spreading. Using the generalized Aubry–André–Harper model, we investigate how the extended, critical, and localized nature of Hamiltonian eigenstates affects the resulting dynamics at fixed dissipation. While the mean-square displacement barely distinguishes these regimes, the ratio of coherent to incoherent currents reveals distinct transport regimes and connects the enhancement of diffusion to quantum coherence. Deviations from its correspondence with the squared Péclet number reveal the contribution of longer-range coherences. Remarkably, dissipation reorganizes coherence in a way that is not directly inherited from the Hamiltonian eigenstates, with critical states sustaining the largest coherence extent. Our results establish spatially resolved coherence as a key diagnostic of quantum active dynamics.
\end{abstract}

\maketitle
\textbf{\textit{Introduction}} ---
Active matter encompasses systems in which microscopic mechanisms of energy conversion sustain motion beyond ordinary thermal diffusion~\cite{bechinger2016active, burgardt2026quantum,
shaebani2020computational, ramaswamy2017active, vrugt2026colloquium, ramaswamy2010mechainics, fodor2016how}. This concept has been recently extended to the quantum domain, identifying signatures of active behavior in a variety of open quantum systems~\cite{penner2025heat, gipouloux2026active, vrugt2026colloquium,
antonov2025engineering, romano2026self, steiner2026active,
antonov2026modeling, khasseh2025active, lee2026anomalous, assent2026quantum, brunner2026quantum, yamagishi2024proposal}.
A central question in this context is how genuinely quantum features contribute to the emergence and properties of active motion. To address this issue, rather than starting from coherent quantum dynamics and investigating how environmental couplings induce activity, we take the complementary perspective, and examine how a classical stochastic dynamics changes upon introducing quantum coherence. 

Consider a particle undergoing both coherent and incoherent hopping on a lattice. In the absence of Hamiltonian dynamics, the particle would perform a classical random walk. Coherent hopping qualitatively modifies its spreading, giving rise to an intermediate superdiffusive regime followed by enhanced diffusion~\cite{gipouloux2026active}. 
Thus, the microscopic mechanism for active-like motion is provided by  quantum coherence~\cite{streltsov2017colloquium,baumgratz2014quantifying,aditya2026coherence, palamara2026revealing, xi2015quantum, francica2019role, francica2020quantum,perciavalle2026quantum}.

Indeed, much as
self-propulsion does on top of Brownian motion~\cite{bechinger2016active},
with dissipation setting its persistence time,
the Hamiltonian generation of coherence on top of an otherwise dissipative dynamics 
induces an active-like behavior, with a diffusion enhancement that reflects precisely the amount of coherence generated.

However, quantum coherence enjoys a spatial structure shaped by the
eigenstates of the underlying Hamiltonian, which can range from extended to
critical to localized~\cite{han1994critical, wang2021many, roy2025many,
anderson1958absence, nandkishore2015many, luitz2014universal,
duthie2022anomalous, chiaracane2021quantum, guo2024multiple, settino2020emergence}. To show the relevance of its spatial distribution, we exploit an exact hierarchy of equations of motion organized according to the range of coherences. 
The hierarchy is dictated by the dynamics itself: the mean-square displacement (MSD)
follows from the populations, which are driven by currents; currents are set
by nearest-neighbor coherences, which are in turn fed by coherences at larger
distances. This structure naturally identifies three levels at which the dynamics can be characterized: particle spreading, the relative contribution of coherent and incoherent transport, and the spatial extent of coherence. 
Each level carries information that cannot be retrieved from those above, as we are going from populations, which are only the classical shadow of a quantum state, towards genuinely quantum features residing in its coherences.

\begin{figure}[!t]
\centering
\includegraphics[width=0.9\columnwidth]{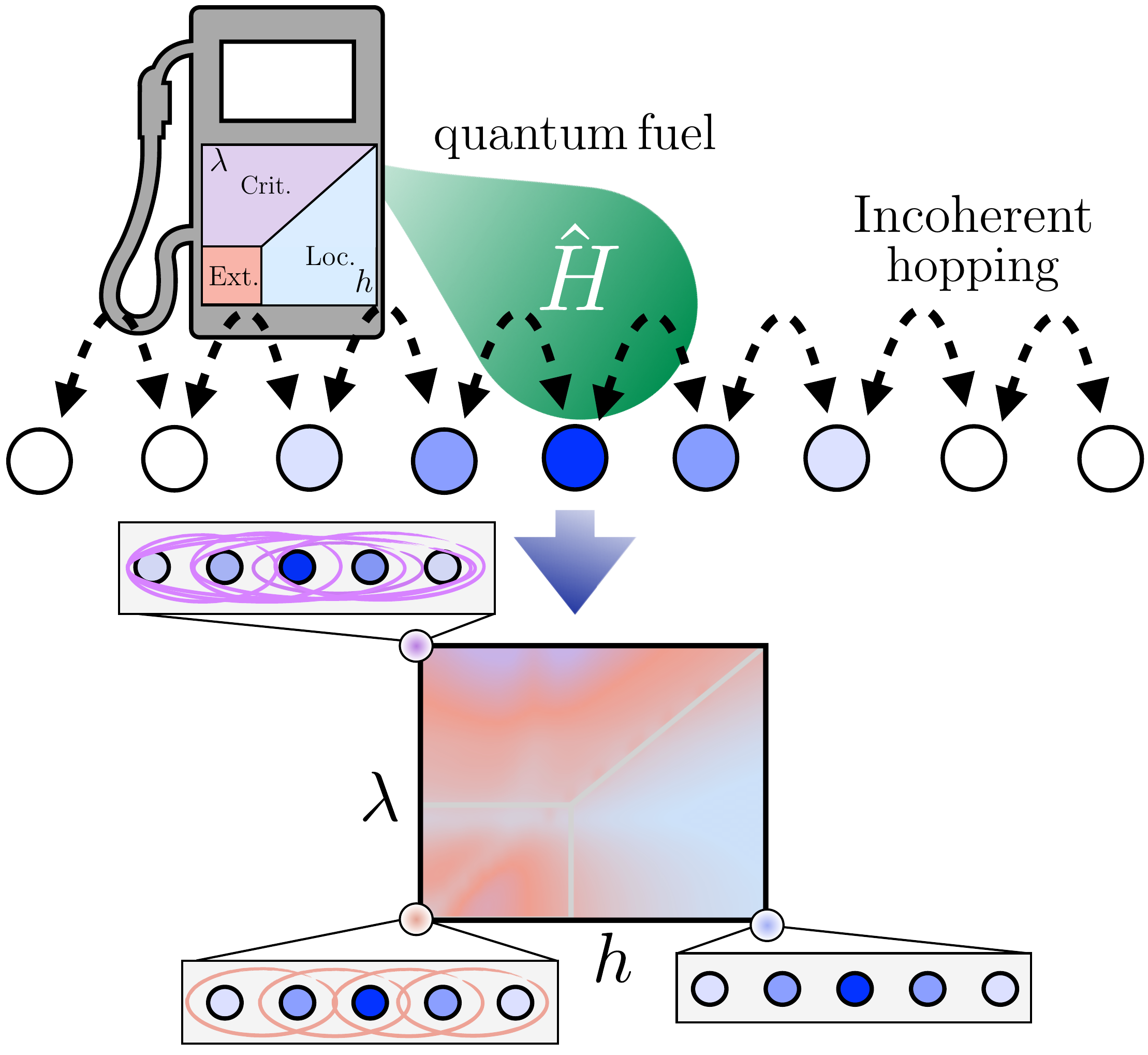}
\caption{Coherent hopping generated by the Hamiltonian $\hat H$ is interpreted as ``fuel'' for active matter when introduced into an otherwise classical stochastic hopping process. The spatial coherence properties of the Hamiltonian eigenstates are controlled by tuning the system across extended (Ext.), critical (Crit.), and localized (Loc.) phases. The interplay between coherent and incoherent hopping gives rise to active-like dynamics and reorganizes and reshapes the spatial distribution of coherence, as illustrated in the lower panel.}
\label{fig:sketch}
\end{figure}
To investigate the interplay between these levels of description, we consider the generalized Aubry–André–Harper  Hamiltonian (GAAH)~\cite{han1994critical, li2023observation, wang2021many, roy2025many, ahmed2025supervised, qi2023multiple, bhakuni2024noise}, including dissipation to model incoherent hopping, see Fig.~(\ref{fig:sketch}). In particular, the Hamiltonian eigenstates can be tuned between extended, critical, and localized phases, thus providing different spatial structures of coherence. 

We find that the simultaneous presence of both coherent and dissipative hopping reshapes the
phase diagram of $\hat H$, as revealed beyond the first level of
the hierarchy. At the top, in fact, the MSD  is only weakly sensitive to the spatial structures; it displays the same sequence of regimes in
all of the phases, while coherence only sets the extent of superdiffusion. One level down, 
we may ascertain whether transport is dominated by
quantum or classical currents. Their ratio broadly follows the phase diagram, with sharp
transitions changed into smooth crossovers, and quantum-dominated transport extending
well into the localized phase. A reshaping of the phase diagram is fully visible in the
longer-range coherences: their interplay with dissipation gives rise to  structures 
within a single phase, displayng features that are absent for the closed system.

These results showcase our main message, that quantum active dynamics cannot be fully characterized in terms of particle spreading or inferred from the properties of the underlying Hamiltonian alone. In fact, the spreading is
the standard way of characterizing classical active systems; but it
cannot help establishing the nature of activity.  Its quantum
origin and its structure require observables that explicitly resolve quantum coherence. 
Two-point correlators, routinely measured, e.g.,  in Rydberg and cold-atom
platforms~\cite{bornet2024enhancing, perciavalle2023controlled,
chen2023continuous, ardila2018measuring}, are therefore the minimal
measurement to identify coherence induced activity.

\textbf{\textit{Quantum Active-like motion}} ---
We consider a single fermion on a chain where coherent Hamiltonian hopping competes
with incoherent, stochastic hopping between neighboring sites. The dynamics
follows the Lindblad equation~\cite{fazio2025many,gipouloux2026active,passarelli2026range}
\begin{equation}
\partial_t\rho
=-i[\hat H,\rho]
+\frac{\Gamma}{2}\sum_j\Big(\mathcal{D}[\hat c_j^\dagger\hat c_{j+1}]
+\mathcal{D}[\hat c_{j+1}^\dagger\hat c_j]\Big)\rho ,
\label{eq:ME}
\end{equation}
where $\mathcal{D}[\hat{\mathcal O}]\rho=\hat{\mathcal O}\rho\hat{\mathcal O}^\dagger
-\frac{1}{2}\{\hat{\mathcal O}^\dagger\hat{\mathcal O},\rho\}$. The dissipators
make the particle hop stochastically at rate $\Gamma/2$ in each direction, while
$\hat{H}=\sum_{j=1}^{L-1}J_j(\hat{c}_j^\dagger\hat{c}_{j+1}+\mathrm{H.c.})
+\sum_{j=1}^L h_j\hat{n}_j$ describes coherent nearest-neighbor hopping in an
on-site potential, with $J_j$ and $h_j$ specified below.

For a particle initially at site $j_0$, the mean-square displacement,
$\mathrm{MSD}(t)=\sum_j(j-j_0)^2\langle\hat n_j(t)\rangle$, has the hallmark
of activity~\cite{gipouloux2026active}. For the homogeneous chain,
$J_j=J$ and $h_j=0$, it is diffusive,
$\mathrm{MSD}\simeq\Gamma t$, for $t\ll t_{\rm sup}\equiv\Gamma/J^2$,
superdiffusive, $\mathrm{MSD}\sim t^\alpha$ with $1<\alpha\lesssim2$, at
intermediate times, and enhanced-diffusive,
$\mathrm{MSD}\simeq\Gamma(1+\mathrm{Pe}^2)t$, for $t\gg t_{\rm enh}\equiv1/\Gamma$,
with the P\'eclet number $\mathrm{Pe}$ measuring the enhancement. This
sequence alone, however, does not prove that the activity is quantum: a
classical process could reproduce the
same curve~\cite{bechinger2016active}.

The reason is that the MSD is a function of the populations
$\langle\hat n_j\rangle$ alone, so its dependence on the quantumness of the
dynamics is hidden. This dependence is made explicit by an exact hierarchy.
By defining $\rho^{(r)} \equiv \rho_{ij}$ with $r=|i-j|$ 
the distance between the two sites, Eq.~\eqref{eq:ME} yields~\cite{perciavalle2026companionLOng}
\begin{equation}
\partial_t\rho^{(r)}
=\mathcal F_r\big[\rho^{(r-1)},\rho^{(r)},\rho^{(r+1)}\big],
\qquad r=0,\dots,L-1.
\label{eq:hierarchy}
\end{equation}
Hopping couples each distance only to $r\pm1$, while dissipation and the
on-site potential act within each distance. The populations ($r=0$) are
driven by the currents, set by nearest-neighbor coherences ($r=1$), which
are in turn fed by longer-range coherences ($r\geq2$). The MSD thus
depends on coherence only indirectly, through every level above it.
Characterizing quantum activity requires probing each level: we use the
MSD for $r=0$, and introduce one observable for $r=1$ and one for
$r\geq2$.

\emph{Level $r=1$: currents.}
The continuity equation splits the particle
current into
$\hat{\mathcal I}_j=\hat{\mathcal I}^{\rm coh}_j+\hat{\mathcal I}^{\rm inc}_j$:
a coherent part,
$\hat{\mathcal I}_j^{\rm coh}=iJ_j(\hat c_j^\dagger\hat c_{j+1}-\mathrm{H.c.})$,
carried by nearest-neighbor coherences, and an incoherent part,
$\hat{\mathcal I}_j^{\rm inc}=\frac{\Gamma}{2}(\hat n_j-\hat n_{j+1})$,
carried by the stochastic redistribution of
populations~\cite{gipouloux2026active}. 
Through them we define the quantum-to-classical current ratio (QCCR),
\begin{equation}
\mathcal R(t)=\frac{\mathcal I^{\rm coh}(t)}{\mathcal I^{\rm inc}(t)},\qquad
\mathcal I^{\rm coh(inc)}(t)=\sum_j\big|\langle\hat{\mathcal I}^{\rm coh(inc)}_j(t)\rangle\big|,
\end{equation}
with $\mathcal R\gg1$ ($\ll1$) for fully coherent (incoherent)
transport. The MSD tells how fast the particle spreads; the QCCR tells
what makes it spread. Truncating Eq.~\eqref{eq:hierarchy} at this level,
$\rho^{(r\geq2)}=0$, links the two currents of the homogeneous chain through
a memory relation~\cite{perciavalle2026companionLOng}, which, summed over
bonds, gives
\begin{equation}
\mathcal I^{\rm coh}(t)\leq\frac{4J^2}{\Gamma}\int_0^t dt'\,e^{-\Gamma(t-t')}\,\mathcal I^{\rm inc}(t'),
\label{eq:memory}
\end{equation}
where the inequality follows from the triangle inequality. Coherent
transport is fed by the history of incoherent transport, through a memory
kernel of width $1/\Gamma$: the Hamiltonian, 
sustains a coherent
channel whose persistence time is set by dissipation.
Two results follow. At short times, $\mathcal R(t)$ grows linearly and
crosses unity at $t_{\rm cross}\simeq\Gamma/4J^2=t_{\rm sup}/4$: quantumness
takes over transport well before the MSD registers it, and superdiffusion
is its delayed consequence. At long times, $\Gamma t_*\gg1$, the memory
relation gives
$\mathcal I^{\rm coh}_*\simeq(4J^2/\Gamma^2)\,\mathcal I^{\rm inc}_*$~\cite{gipouloux2026active,perciavalle2026companionLOng},
with $f_*\equiv f(t_*)$, hence
$\mathcal R_*\simeq4J^2/\Gamma^2=\mathrm{Pe}^2$ and
$D_{\rm enh}=\frac{\Gamma}{2}(1+\mathcal R_*)$. The enhancement of
diffusion equals the weight of the coherent channel: coherence does not
merely accompany enhanced diffusion, it causes it. Just as self-propulsion
adds directed motion on top of Brownian motion, with $\mathrm{Pe}$
measuring its weight~\cite{bechinger2016active}, coherent hopping adds a
quantum channel on top of stochastic hopping, and the QCCR measures its
weight.

\emph{Level $r\geq2$: coherence extent.}
$\mathcal R_*\simeq\mathrm{Pe}^2$ relies only on the nearest-neighbor
truncation, not on homogeneity: within it, the correspondence holds for
arbitrary $J_j$ and $h_j$, with the populations mapping onto a
random-barrier walk and
$\mathcal R_*\simeq\mathrm{Pe}^2=\langle\!\langle(1+\kappa_j)^{-1}\rangle\!\rangle^{-1}-1$,
where $\kappa_j=4J_j^2/[\Gamma^2+(h_j-h_{j+1})^2]$ and
$\langle\!\langle\cdot\rangle\!\rangle$ averages over
bonds~\cite{perciavalle2026companionLOng}. Deviations from this correspondence at long times are therefore
a direct signature of longer-range coherences feeding back into the
currents. To probe them, we introduce the $r$-distance coherence
fraction
\begin{equation}
\mathcal C(r)=\frac{g(r)}{\sum_{r'=1}^{L-1}g(r')},
\qquad
g(r)=\frac{1}{2(L-r)}\sum_{|i-j|=r}|\rho_{ij}|,
\label{eq:coh_fraction}
\end{equation}
i.e., the relative weight of the coherence between a typical pair of sites at distance $r$ (normalized per pair so as not to favor short distances), 
and its participation ratio, the coherence extent
$\xi=\big[\sum_{r=1}^{L-1}\mathcal C(r)^2\big]^{-1}$~\cite{ luitz2014universal, evers2008anderson}.
$\xi$ counts the distances over which coherence effectively extends, from
$\xi\simeq1$ (nearest-neighbor coherence only) to $\xi\simeq L-1$ (fully
delocalized coherence, $|\rho_{ij}|=1/L$). It measures exactly what
population-based observables miss: how far coherence reaches away from the
diagonal of $\rho$.

\textbf{\textit{Spatially tuning coherence: the generalized Aubry--Andr\'e--Harper model}} ---
We have argued that coherence is the true source of active-like motion. We now ask how its
\emph{spatial structure} matters, that is, whether the way coherence is
generated and spread across the system shapes the resulting dynamics. This
requires a Hamiltonian whose eigenstates can be tuned from extended to
localized in a controlled way. The generalized Aubry--Andr\'e--Harper (GAAH)
model~\cite{han1994critical,li2023observation,wang2021many,roy2025many,ahmed2025supervised}
does exactly this. It is obtained from $\hat H$ with quasiperiodic hopping
$J_j=|J+\lambda\cos(2\pi\beta(j+\tfrac12)+\phi)|$ and potential
$h_j=h\cos(2\pi\beta j+\phi)$, where $\beta=(\sqrt5-1)/2$. Already for a
single particle, it hosts extended, critical (multifractal), and localized
phases~\cite{han1994critical,li2023observation,roy2025many,Li2022generalized,ahmed2025supervised,zhou2026fundamental,huang2026observation}.
These are separated by sharp transitions at $h_c=2J$ for $\lambda<J$, at
$\lambda_c=J$ for $h<2J$, and along $\lambda=h/2$ for
$h,\lambda>J$~\cite{li2023observation,roy2025many,wang2021many}.
We set $J=1$. The three phases provide three structurally distinct type of states:
eigenstates spread over the whole chain, multifractal eigenstates, and
eigenstates confined to a few sites. The GAAH model lets us compare them
within a single framework and at fixed dissipation.

This setup is within reach of current experiments. The GAAH model has been
realized with ultracold atoms and superconducting
circuits~\cite{li2023observation,huang2026observation}. Dissipative hopping
can be engineered in cold-atom and Rydberg
platforms~\cite{haga2021liouvillian,schempp2015correlated}. All our
diagnostics require only site populations and two-point correlators
$\langle\hat c_i^\dagger\hat c_j\rangle$. In the single-particle sector these
reduce to the spin correlators $\langle\hat\sigma^+_i\hat\sigma^-_j\rangle$,
which are measured routinely in Rydberg arrays and ultracold-atom
experiments~\cite{bornet2024enhancing,perciavalle2023controlled,chen2023continuous,ardila2018measuring}.

\textbf{\textit{Transient Active Dynamics}} ---
\begin{figure}[!t]
\centering
\includegraphics[width=1\columnwidth]{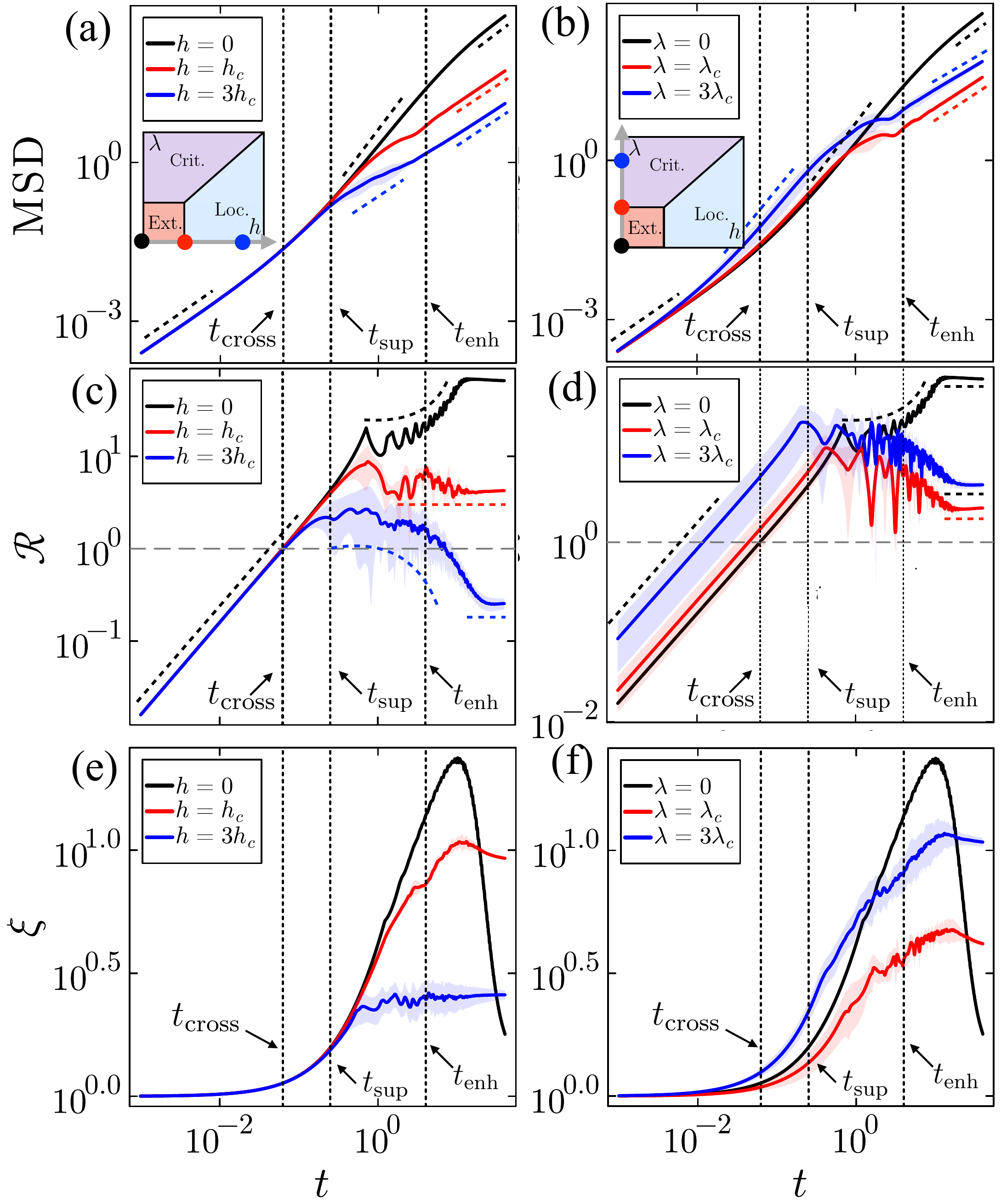}
\caption{Particle spreading, QCCR, and coherence extent in paradigmatic
regimes, across the extended--localized transition at $\lambda=0$
[(a),(c),(e)] and the extended--critical transition at $h=0$
[(b),(d),(f)]. (a),(b): MSD; (c),(d): QCCR; (e),(f): coherence extent
$\xi$. For $h=\lambda=0$.
Vertical dotted lines mark $t_{\rm cross}$,
$t_{\rm sup}$, $t_{\rm enh}$. The insets in (a),(b) show the phase diagram
of the closed system in the $(h,\lambda)$ plane, with dots marking the
parameters analyzed. $L=161$, $N_\phi=100$, $\Gamma=1/4$.}
\label{fig:dyn}
\end{figure}
\begin{figure*}[!t]
\centering
\includegraphics[width=\linewidth]{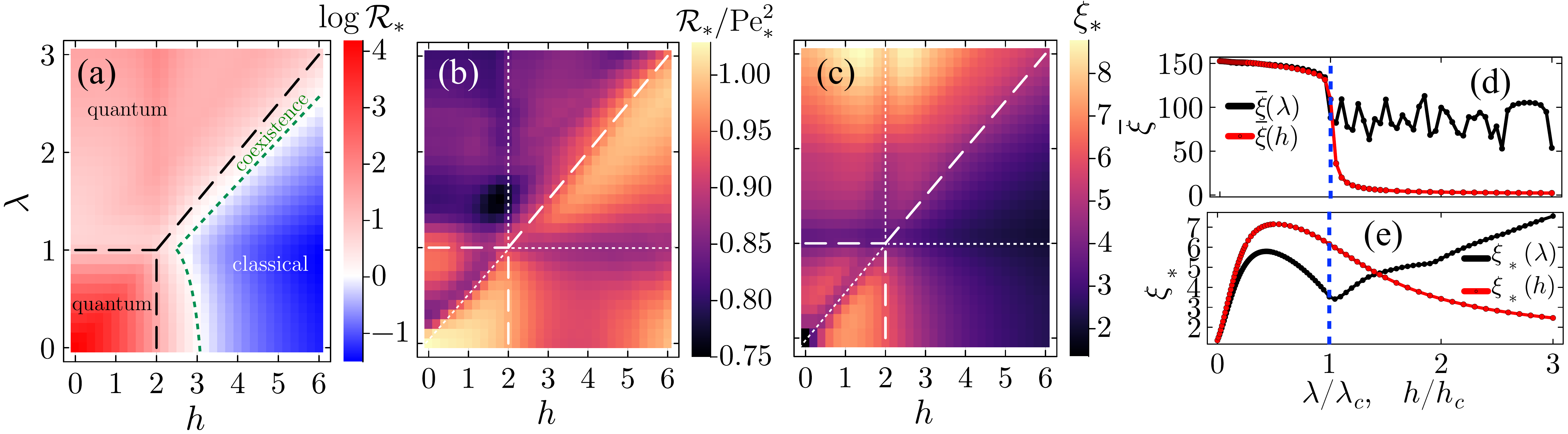}
\caption{Enhanced diffusive regime, $t_*=10/\Gamma$. (a)–(c): QCCR, correspondence $\mathcal{R}_*/\mathrm{Pe}_*^2$, and coherence extent $\xi_*$ in Hamiltonian parameter space (dashed: Hamiltonian critical lines; dotted in (b),(c): their continuations, along which the maps change sharply). (d),(e): coherence extent averaged over the eigenstates of the bare Hamiltonian, $\bar\xi$ (d), and evaluated at $t_*$, $\xi_*$ (e); red: $\xi(h)$ at $\lambda=0$; black: $\xi(\lambda)$ at $h=0$. $\phi=0$, $L=161$.}
\label{fig:long}
\end{figure*}
We scan $h$ at $\lambda=0$ and $\lambda$ at $h=0$, crossing the extended,
critical, and localized phases, and average over $N_\phi=100$ random phases
$\phi$ [Fig.~\ref{fig:dyn}]. At the top of the hierarchy, the MSD barely
distinguishes the phases. Along $h$ [Fig.~\ref{fig:dyn}(a)], the curves
coincide up to $t_{\rm sup}$. Beyond this time, the superdiffusive window
shrinks as localization sets in and incoherent hopping takes over. In the
localized phase, only a weak, $\phi$-dependent shoulder survives for
$t_{\rm sup}\lesssim t\lesssim t_{\rm enh}$. Along $\lambda$
[Fig.~\ref{fig:dyn}(b)], the modulation enters the hopping amplitudes and
renormalizes the timescales themselves: at $\lambda=3\lambda_c$,
superdiffusion sets in markedly earlier than at $\lambda=0$. In all phases,
the long-time MSD is diffusive, with an enhancement that is non-monotonic in
$\lambda$. Quantum coherence changes how fast and how far the particle spreads, but
not the sequence of spreading regimes.

One level down, the QCCR resolves what the MSD cannot
[Fig.~\ref{fig:dyn}(c),(d)]. At short times, $\mathcal R(t)$ grows linearly
in all phases. It crosses unity at $t_{\rm cross}$ in the extended and
localized phases, and earlier in the critical phase, where coherent
transport builds up faster. Beyond this point, the phases split. In the
extended phase, $\mathcal R(t)$ keeps growing as
polylogarithmically before saturating at
$\mathcal R_*\simeq\mathrm{Pe}_*^2$: coherent transport builds up
progressively and survives asymptotically. In the critical phase,
$\mathcal R(t)$ develops a broad, strongly fluctuating maximum, a long
window of coherence-dominated transport. Its long-time value is
non-monotonic in $\lambda$ and does not follow the average hopping: at
$\lambda=3\lambda_c$ it is substantially smaller than at $\lambda=0$,
despite the larger average hopping, reflecting the multifractal
eigenstates. In the localized phase, coherence is not immediately
suppressed. $\mathcal R(t)$ first rises above
unity, then slowly decays and falls below it shortly after $t_{\rm enh}$.
This window shows strong $\phi$-dependent fluctuations as residual
coherent dynamics competes with dissipation. In all phases,
$\mathcal R(t)$ relaxes only beyond $t_{\rm enh}$, the common dissipative
timescale.

Hence, although the long-time MSD is diffusive in all phases, the QCCR yields
a transport classification invisible to it~\cite{perciavalle2026companionLOng}:
\begin{equation}
\text{Transport}:
\begin{cases}
\mathrm{Inc.}\rightarrow\mathrm{Coh.}, & \text{extended, critical},\\
\mathrm{Inc.}\rightarrow\mathrm{Coh.}\rightarrow\mathrm{Inc.}, & \text{localized}.
\end{cases}
\end{equation}
Localization does not eliminate coherent transport but makes it
transient. This transient window requires weak enough dissipation and
disappears for $h\gg h_c$, where eigenstates are nearly site-localized
and residual coherence is almost
extinguished~\cite{perciavalle2026companionLOng}. Yet the QCCR is built
from nearest-neighbor coherences alone: it tells how much of the
transport is quantum, not how coherence is organized at larger distances.

This is resolved at the bottom of the hierarchy by $\xi(t)$
[Fig.~\ref{fig:dyn}(e),(f)]. In the homogeneous chain, $\xi$ grows
throughout the superdiffusive window, peaks after $t_{\rm enh}$, and then
collapses to nearest neighbors. This coherence-extent barrier is a
hallmark of the homogeneous extended phase, and is broken as soon as
quasiperiodicity is introduced: $\xi$ relaxes over several sites, most
extensively in the critical phase, while deep in the localized phase it
saturates early, at a few sites, with no barrier. Coherent transport and
coherence extent are thus ordered differently: the homogeneous chain
carries the most coherent transport, yet ends with the least extended
coherence in the enhanced-diffusive regime $t \gg t_{\rm enh}$.

\textbf{\textit{Coherence in the enhanced-diffusive regime}} ---
At long times, $t_*=10/\Gamma$, the dynamics reaches enhanced diffusion,
$\mathrm{MSD}\simeq\Gamma(1+\mathrm{Pe}^2)t$, everywhere in the phase diagram.
Fluctuations in $\phi$ are weak here, so we fix $\phi=0$ and analyze the role of coherence going down the
hierarchy level by level.

At the top, the MSD sees a single diffusive state. One level down, the QCCR
resolves three transport regimes [Fig.~\ref{fig:long}(a)], and its map is
reminiscent of the Hamiltonian phase
diagram. Enhanced diffusion is quantum
($\mathcal R_*>1$) in the extended and critical phases and classical
($\mathcal R_*<1$) deep in the localized phase. Dissipation, however, turns
the sharp transitions of the closed system into smooth crossovers. The
quantum and classical regimes are separated by a
\emph{quantum--classical coexistence} line, $\mathcal R_*\simeq1$, which is
shifted inside the localized phase, away from the closed-system transition.
The residual coherence of localized eigenstates thus sustains
quantum-dominated diffusion well beyond the transition point.

The QCCR, however, measures only how \emph{much} transport is coherent, and
its link to spreading, $\mathcal R_*\simeq\mathrm{Pe}_*^2$, assumes that
coherence does not extend beyond nearest neighbors. We test this link by
extracting the P\'eclet number directly from the MSD,
$\mathrm{Pe}_*^2=\tilde D(t_*)/\Gamma-1$ with $\tilde D(t)=d\,\mathrm{MSD}/dt$
[Fig.~\ref{fig:long}(b)]. The relation holds near $h=\lambda=0$ and in the
localized phase, but fails by up to $\sim25\%$ in the critical phase. These
deviations signal longer-range coherences feeding back into the currents,
which the lowest level of the hierarchy makes visible. The coherence extent $\xi_*$ [Fig.~\ref{fig:long}(c)] shows how
coherence is \emph{organized}, and draws a different map from the QCCR.
At $h=\lambda=0$, the coherence is fully extended, yet the collapse of the
coherence-extent barrier leaves $\xi_*\simeq1$. Any quasiperiodic
modulation, however weak, breaks the barrier dynamical structure, and coherence survives over
several sites. In the localized phase, instead, $\xi_*$ remains small
because the Hamiltonian generates coherence over only a few sites. $\xi_*$
varies strongly within a single phase and changes sharply along the
continuations of the critical lines (dotted), where the closed system has
no transition. These are also the borders of the P\'eclet--QCCR
correspondence map [Fig.~\ref{fig:long}(b)], and $\xi_*$ is largest in
the critical phase, where the correspondence fails most.

A natural question is wether this structure is inherited from the eigenstates. This is true only in the localized
phase, where both span a few sites [Fig.~\ref{fig:long}(d),(e)].
Elsewhere in parameter space, dissipation reshapes coherence
[Fig.~\ref{fig:sketch}]. It
confines it: eigenstates spanning the chain, $\bar\xi\sim L$,
yield $\xi_*$ of a few sites. And it reorganizes it: the most
extended eigenstates ($h=\lambda=0$) end with the least coherence,
critical ones with the most, and the jumps of $\bar\xi$ at the
transitions are smoothed out.

As a final note, we emphasize that the picture is truly hierarchical and that the features are not
finite-size effects, as both $\mathcal R(t)$ and $\xi(t)$ depend only weakly on system
size~\cite{perciavalle2026companionLOng}. 

\textbf{\textit{Conclusions and Outlook}} ---
We have investigated how quantum coherence induces and shapes active-like motion in an otherwise classical stochastic process. An exact hierarchy of equations of motion provides a natural framework for separating three complementary aspects of the dynamics: particle spreading, coherent transport, and the spatial organization of coherence. Going beyond spreading is essential to reveal the true role of quantum coherence  within activity and to characterize it.

The ratio of coherent to incoherent currents quantifies the contribution of Hamiltonian dynamics to transport. Within the nearest-neighbor coherence approximation, this ratio is directly related to the squared Péclet number, connecting the enhancement of diffusion to the coherent contribution to particle currents, a relation that holds for any nearest-neighbor Hamiltonian with an inhomogeneous on-site field~\cite{perciavalle2026companionLOng}. Deviations from this correspondence reveal the feedback of longer-range coherences on transport.

By tuning the underlying Hamiltonian across extended, critical, and localized phases at fixed dissipation, we have shown that the properties of the resulting dynamical state cannot be directly inferred from those of the Hamiltonian eigenstates. While the current ratio retains signatures of the closed-system phase diagram, the spatial distribution of coherence exhibits a qualitatively different structure. In particular, critical eigenstates sustain more extended coherences than the homogeneous chain, despite the latter exhibiting a larger coherent contribution to transport. These differences are imprinted in distinct density-matrix structures that persist in the enhanced-diffusive regime and are robust against system size~\cite{perciavalle2026companionLOng}.

These findings demonstrate that the magnitude of coherent transport and the spatial extent of coherence characterize distinct aspects of quantum active dynamics. Particle spreading alone is insufficient to resolve either of them. Spatially resolved two-point correlators provide a direct means of probing these features. They are accessible in Rydberg arrays and ultracold-atom platforms \cite{bornet2024enhancing,
perciavalle2023controlled, chen2023continuous, ardila2018measuring,
schempp2015correlated},  which enjoy also the possibility to explore
how coherence built over large distances competes with quasiperiodic
modulation and dissipation~\cite{defenu2023long, defenu2024out,
passarelli2026range, russomanno2023entanglement}.

Our framework can be extended to investigate how longer-range hopping and many-body interactions affect the emergence of quantum active dynamics. More broadly, identifying
the role of Hamiltonian induced coherence suggests a possible
resource-theoretic perspective on active
dynamics~\cite{chitambar2019quantum, bhakuni2026quantum,
falcao2025nonstabilizerness}.

\textbf{\textit{Acknowledgements}} --- 
We thank G. Fava, G. Tucci, O. La Rocca, A. Palamara for fruitful discussions. Numerical simulations were performed using the Julia package
\texttt{QuantumToolbox.jl}~\cite{QuantumToolbox.jl2025}. The authors used Claude (Anthropic) and ChatGPT (OpenAI) for assistance with coding and editing. All outputs were checked by the authors, who takes full responsibility for the content. We acknowledge support from PNRR MUR Project No. PE0000023-NQSTI through the secondary project ThAnQ.

\end{document}